\documentclass[twoside,fleqn]{article}
\usepackage{epsf}
\usepackage{espcrc2}
\usepackage{latexsym}

\newcommand{\LL}{\left\langle}
\newcommand{\RR}{\right\rangle}

\newcommand{\BE}{\begin{equation}}
\newcommand{\EE}{\end{equation}}
\newcommand{\BEA}{\begin{eqnarray}}
\newcommand{\EEA}{\end{eqnarray}}
\newcommand{\EL}{\nonumber\\}
\newcommand{\Dslash}{/ \!\!\!\! D}

\newcommand{\psl}{/ \!\!\!\! p}

\newcommand{\etal}{{\em et al.\ }}

\newcommand{\ie}{{\em i.e.\ }}

\newcommand{\EQ}[1]{Eq. (\ref{#1})}
\newcommand{\REF}[1]{Ref.~\cite{#1}}
\newcommand{\FIG}[1]{Fig. (\ref{#1})}

\newcommand{\gbeta}{\beta}
\newcommand{\pbp}{\bar\psi\psi}

\def\simge{%  ``greater than about'' symbol
    \mathrel{\rlap{\raise 0.511ex
        \hbox{$>$}}{\lower 0.511ex \hbox{$\sim$}}}}
\def\simle{%  ``less than about'' symbol
    \mathrel{\rlap{\raise 0.511ex
        \hbox{$<$}}{\lower 0.511ex \hbox{$\sim$}}}}

\newcommand{\AmS}{{\protect\the\textfont2
  A\kern-.1667em\lower.5ex\hbox{M}\kern-.125emS}}

\title{Domain wall fermions in vector gauge theories}

\author{T. Blum\address{Physics Department, Bldg. 510A,
	Brookhaven National Lab\\Upton, NY 11973-5000, USA}}
       
\begin{document}

\begin{abstract}
I review domain wall fermions in vector gauge theories. 
Following a brief introduction, the 
status of lattice calculations using domain wall fermions is
presented. I focus on results from QCD, including
the light quark masses and spectrum,
weak matrix elements, the $n_f=2$ finite temperature phase transition, and
topology and zero modes and conclude with topics for future study.
\end{abstract}

\maketitle

\section{INTRODUCTION}
Several years ago Kaplan had the idea for a lattice discretization of
the Dirac operator that preserves chiral symmetry at any
lattice spacing\cite{KAPLAN}. This miracle is performed by adding an {\it odd}
infinite dimension to the usual {\it even} dimensional spacetime. The mass of the fermion is
given the shape of a domain wall in this extra dimension. The mass is
large, positive on one side of the defect, negative on the other. 
Chiral zero modes then arise
naturally on the wall where the mass is zero. 
If the extra dimension is periodic, an anti-domain wall
appears at the other end of the lattice with a zero mode of the opposite chirality. These two
chiral zero modes are coupled to the same (4d) gauge field to simulate a vector gauge theory,
like QCD. The chiral symmetry is manifest: the left and right handed fermions 
have been separated globally in the extra dimension, so they may be rotated independently.

This realization of chiral zero modes was already known to occur in the continuum, 
for instance, by Callan and Harvey
who studied topological defects embedded in higher dimensions\cite{CandH}. There and in
Kaplan's original study one may suppose that the extra 5th dimension is physical: our 4d
spacetime is bound to the wall at low energies, but at high enough energies the 5th
dimension becomes perceivable. It is interesting to note such ideas have recently been
proposed as a natural mechanism for the spontaneous breaking of supersymmetry\cite{SHIFMAN}.
Loosely speaking, domain walls are D-branes, and similar models for realistic 
``compactified" universes have recently been proposed(for example, see \REF{SUND}).

Shortly after Kaplan's original proposal and also motivated by an independent proposal
by Frolov and Slavnov\cite{FROLOV}, Naryanan and Neuberger transformed the extra
dimension into an infinite flavor space of heavy regulator fields\cite{NN1}. 
Using a transfer matrix formalism, the chiral determinant is given as the overlap of
the ground states of two Hamiltonians(hence the name, the overlap formulation, or the overlap,
for short). The overlap and Kaplan's domain wall fermions(DWF) are equivalent in the particular
case where the extra dimension is infinite.
The overlap provides an elegant framework for studying the Index theorem on the lattice.
It is the starting point for a new lattice Dirac operator that also maintains
chiral symmetry on the lattice\cite{NEU1}(see below). 
An extensive review of the overlap is given in \REF{NN2}.

The original proposal by Kaplan was intended as a lattice chiral gauge theory. However,
it was soon recognized that the discretization would be useful for lattice 
QCD\cite{SHAMIR1} since conventional methods explicitly break chiral symmetry.
Recent numerical studies have shown
the actual extent of the extra dimension can be modest, 10-20 sites for
simulations of quenched QCD, and still preserve the chiral symmetry to a high degree of 
accuracy\cite{BS1,BS2,BS3}. Further, DWF are accurate to order $O(a^2)$ 
since no (mass) dimension
five operators that are also chirally symmetric 
exist to cancel $O(a)$ errors\cite{NEU2,BS2}. Indeed, the initial numerical results
indicate improved scaling. This scaling (if it holds up after further scrutiny) offsets
the cost of the extra dimension. 

For a finite extra dimension with $N_s$ sites,
the leading discretization errors are expected
to be $O(a)\exp(-C N_s)$ where $C$ 
is the rate of exponential fall off in the extra dimension.
This has been verified at one loop in perturbation theory for QCD\cite{AOKI}.
In particular, the authors of \REF{AOKI} calculate the one loop fermion self-energy and
find that the mass of the light mode remains exponentially small.
Even after accounting for exponentially small 
terms in the free propagator that were neglected
in \REF{AOKI}, \REF{NEU-KIK} also concludes that the 
exponential suppression stays intact after one loop radiative corrections.

At this meeting impressive new results were presented for lattice QCD using DWF,
including the first dynamical simulations\cite{VRANAS3}. 
After a brief description of the method,
I focus on the lattice QCD results.

%A recent review of DWF is also given in \REF{JANSEN}.
\section{BOUNDARY FERMION VARIANT\cite{SHAMIR1}}
Shortly after Kaplan's discovery, Shamir reformulated DWF
using a lattice in the extra dimension that is half as big.
The boundaries of the 5th dimension are explicitly coupled by a small
parameter, $-m (m>0)$. We will see that $m$ is proportional to the $4d$ quark mass and
receives only multiplicative renormalizations in the limit $N_s\to\infty$. The chiral
limit of DWF is thus $N_s\to\infty$ and then $m\to 0$. These 
features make Shamir's variant of DWF useful for numerical simulations.

The action for DWF is essentially an ordinary 5d Wilson fermion action:
\BEA
S_{dwf} &=&-\sum_{x,y,s,s^\prime}\bar\psi \left(\Dslash_{x,y}\delta_{s,s^\prime}+
\Dslash_{s,s^\prime}\delta_{x,y}\right) \psi, \EL
\Dslash_{x,y}&=& \frac{1}{2}\sum_{\mu}\left(\right .
(1+\gamma_\mu)U_{x,\mu}\delta_{x+\hat\mu,y}\EL
&+&
(1-\gamma_\mu)U^\dagger_{y,\mu}\delta_{x-\hat\mu,y}\left . \right)+
(M-4)\delta_{x,y}, \EL
\Dslash_{s,s^\prime}&=& 
\begin{array}{ll}
P_R\delta_{1,s^\prime}
-m P_L\delta_{N_s-1,s^\prime}-\delta_{0,s^\prime} \EL%& s=0\EL
P_R\delta_{s+1,s^\prime}+
P_L\delta_{s-1,s^\prime}-\delta_{s,s^\prime}\EL% & 1\le s\le N_s-2\EL
-m P_R\delta_{0,s^\prime}+
P_L\delta_{N_s-2,s^\prime}-\delta_{N_s-1,s^\prime}\EL% & s=N_s-1
\end{array},\\
\label{ACTION}
\EEA
where $M$ is the full 5d mass parameter, $s$ and $s^\prime$ denote the extra dimension, and
$P_{R,L}=(1\pm\gamma_5)/2$.
There are important differences with the ordinary Wilson action, however. The links in the 
$s$ direction are set to unity,
the relative sign between the Wilson term and
the 5d mass term is opposite to the usual convention, and, as mentioned above,
the boundaries in the $s$ direction are coupled with a strength $m$. The boundary
conditions are anti-periodic since the Kaplan model is periodic over $2 N_s$.
Finally, the extra hopping terms have $\gamma_5$'s in place of
$\gamma_\mu$'s; there is no chirality operator in odd dimensions.

In the free theory for $N_s=\infty$ and $p_\mu, m\to 0$
it is straightforward to show that
the mass of the light quark is $m_q=m M (2-M)$ 
and that the singular part of the propagator (on one wall) is
\BEA
G_R(s,s^\prime,p)&=&P_R\frac{M(2-M)}{\psl+m_q} (1-M)^{s+s^\prime}.
\label{free prop}
\EEA
A similar result holds for the other wall.
Aside from the extra factor $M(2-M)(1-M)^{s+s^\prime}$, this is same form found in the continuum. The
free propagator is exponentially suppressed in the extra dimension with a rate that depends
on $M$. For $M=1$ the mode does not penetrate the extra dimension at all. 
Note, the doublers have been removed for the choice $0<M<2$. As $M$ is increased in increments of
two, up to ten, the zero modes disappear from the spectrum and 
new ones appear, depending on which
corners of the Brillouin zone, according to the factor 
$b(p)=1-M+\sum_\mu(1-\cos(p_\mu))$, contribute(see \cite{SHAMIR1} for details).
After $M=10$, no zero
modes exist. The critical value where the light quark first appears is $M_c=0$.
In \REF{VRANAS2}, the leading order corrections to the quark mass for finite $N_s$ were given,
\BEA
m_q=M(2-M)(m + (1-M+O(p^2))^{N_s}).
\label{quark mass}
\EEA
The overlap of the two modes in the extra dimension
has generated an exponentially small quark mass. This mass can be made arbitrarily 
small compared to $m$ by increasing $N_s$.

In the interacting theory $M$ is additively renormalized, just like
ordinary Wilson fermions, $M\to \widetilde{M}=M-M_c$. At one loop, the dominant effect comes
from the tadpole diagram which is diagonal in $s$\cite{AOKI}; 
hence the above free field results are unchanged 
up to a simple shift in $M$. If one uses a reasonable choice for the coupling constant,
then the magnitude of the tadpole contribution agrees nicely with the original
nonperturbative estimate (near quenched $\gbeta\approx 6.0$)
for the optimal value of $M$ which minimizes the overlap of the 
two light modes\cite{BS1,BS2}. A similar shift occurs for the dynamical simulations,
as well\cite{VRANAS3}. \EQ{quark mass} also suggests why we do not expect
the usual ``exceptional" configurations for DWF for {\it even} $N_s$: the contribution
to $m_q$ is always positive.

Four dimensional quark fields are constructed from the five dimensional fields by taking
their chiral projections on the boundaries.
\BEA
q_x &=& \frac{(1+\gamma_5)}{2}\psi_{x,0}+ 
\frac{(1-\gamma_5)}{2}\psi_{x,N_s-1},\EL
\bar q_x &=& \bar\psi_{x,N_s-1}\frac{(1+\gamma_5)}{2}+ 
\bar\psi_{x,0}\frac{(1-\gamma_5)}{2}.
\label{quark ops}
\EEA
These definitions are the simplest choice for interpolating operators that
create and destroy quarks. Other definitions are possible; \ie one may
average over some width around each boundary. 

Operators constructed from the definitions in \EQ{quark ops} satisfy a set of 
axial Ward identities\cite{SHAMIR2}.
\BEA
\Delta_\mu \LL A^a_\mu(x) O(y)\RR &=& 2 m \LL J_5^a(x)O(y)\RR \EL
+ i \LL \delta_A^a O(y)\RR &+& 2 \LL J_{5q}^a(x)O(y)\RR.
\label{cwi}
\EEA
The identities are derived in the usual way, by demanding invariance 
of expectation values under infinitesimal transformations. 
Here, we take advantage of the fact
that the left and right handed modes are globally separated in the fifth
dimension in order to rotate them independently.
The first two terms on the r.h.s appear in the continuum 
while the last term is ``anomalous". For the flavor nonsinglet case it
vanishes in the limit $N_s\to \infty$ before the continuum limit is
taken\cite{SHAMIR2}. Thus, DWF fermions have the full axial symmetry
of the continuum at {\it any} lattice spacing. However, this does not
rule out the possibility that the doublers reappear at strong coupling
in order to trivially satisfy \EQ{cwi}. For the singlet case, the last
term gives rise to the usual axial anomaly\cite{SHAMIR3}.

\section{NUMERICAL RESULTS}

The first numerical works using DWF for vector gauge theories 
were studies of the two dimensional
vector Schwinger model\cite{JASTER,VRANAS1}. Both indicated DWF were practical
and useful for numerical work. An in-depth study of DWF using the two dimensional
vector Schwinger can be found in \cite{VRANAS2}.

The first simulations of QCD using DWF are found in Refs.\cite{BS1,BS2,BS3}. There
it was shown that even modest numbers of sites in the extra dimension were 
sufficient to maintain the desirable chiral properties of DWF. Within the last
year a significant interest in using DWF for QCD has developed, leading to many new
results that were presented at this conference. These are described
below.

\subsection{PCAC Ward identity, zero modes, topology, and finite $N_s$ effects}

The anomalous term in \EQ{cwi} for the PCAC Ward identity ($O=J_5$) can
be used to estimate the effect of finite $N_s$ in simulations. Taking the
ratio of \EQ{cwi} with the pseudoscalar correlator leads to
$2 m + 2 \LL J_{5q}(x)J_5(y)\RR/{\LL J_5(x)J_5(y)\RR}$
on the r.h.s., so, as long as the second term is small compared
to the explicit quark mass $m$, one expects good chiral behavior.

\begin{figure}[hbt]
\vbox{ \epsfxsize=3.0in \epsfbox[0 0 4096 4096]{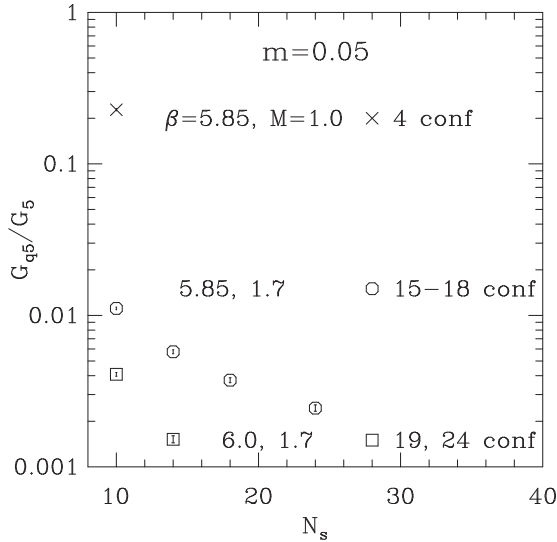} }
\vspace{-.25in}
\caption{The relative anomalous contribution to the PCAC Ward identity.}
\label{anom ratio}
\end{figure}

Results for the relative anomalous contribution are 
shown in \FIG{anom ratio}. 
For $M=1.7$ all the points are at least an order of magnitude less than
the value of $m(0.05)$ used in the simulations, except the point at $N_s=10$
and $\gbeta=5.85$. As shown below, these simulations display good chiral
behavior whereas the pion mass, for example, at $\gbeta=5.85$ and $N_s=10$ 
has a small but non-zero value when extrapolated to $m=0$. Note also that the effects
decrease significantly as the coupling is weakened while the rate appears to 
increase.

The data in \FIG{anom ratio} is plotted on a semi-log plot, so the four points at 
($\gbeta=5.85$, $M=1.7$) should lie on a straight line if the suppression of chiral
symmetry breaking effects is given by a simple exponential. 
However, there is noticeable curvature (but, note that 
the sample of configurations is small).
This curvature may indicate power law suppression or 
exponential suppression with an $N_s$
dependent rate. The latter was observed in the two dimensional 
vector Schwinger model\cite{VRANAS2} and was attributed to gauge
fields having trivial versus nontrivial topology.

The former case may arise from translationally invariant modes in
the 5th dimension which are expected to be the dominant anomalous
contributions to the Ward identities\cite{SHAMIR2}. Such
modes may arise when the 4d Wilson-Dirac operator ($\Dslash_W(-M)$)
supports an exact zero mode(here $-M$ 
refers to supercritical values of the hopping parameter). 
This is because the spectrum of the Hamiltonian
$H_W(-M)\equiv \gamma_5 \Dslash_W(-M)$ defines a transfer matrix 
for propagation in the 5th dimension\cite{NN2},
$T^{N_s}\equiv \exp({-H_W})^{N_s}$.
When $H_W(-M)$ has a zero eigenvalue, $T$ has a unit eigenvalue, and there is no
suppression in the extra dimension. There is a gap in the spectrum below
the $M$ corresponding to $\kappa_c$, which vanishes
at $\kappa_c$. Naively, the gap is expected to reopen above 
this value of $M$\cite{SCRI}.
The spectrum of $H_W(-M)$
has been calculated recently\cite{SCRI,NAR1} on quenched configurations
corresponding to those in \FIG{anom ratio}. It was found that $H_W(-M)$
contains many zeroes, or level crossings, 
arising from lattice instanton-like artifacts, which could close the gap for all 
values of $M$ suitable for DWF simulations, a disaster.

The data for DWF, on the other hand, are not consistent with this scenario.
If there were no suppression, light modes would not be bound to the walls and
the ratio of pseudoscalar densities in \FIG{anom ratio} would be $O(1)$. However, 
we have already seen that this ratio is small for $M=1.7$ which is far from
the $\kappa_c$ region. For $M=1$ which is near this
region at $\gbeta=5.85$, the ratio increases
dramatically, signalling the absence of light modes. This is consistent with
the gap closing at $\kappa_c$, as it should. It may be somewhat confusing that
the existence of light modes for DWF rules them out for Wilson fermions,
and vice-versa.

\begin{figure}[hbt]
\vbox{ \epsfxsize=3.0in \epsfbox{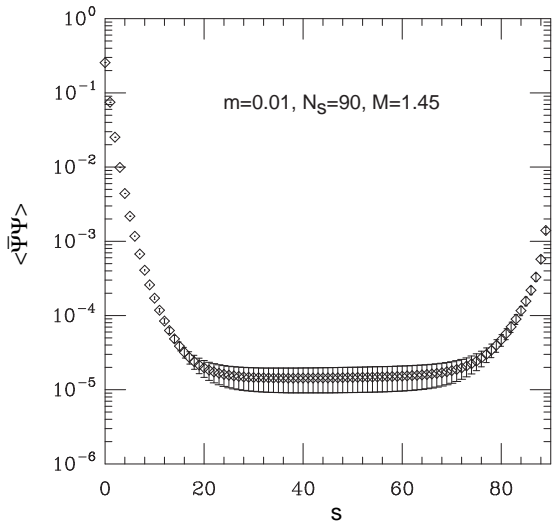} }
\vspace{-.25in}
\caption{$\pbp$ as a function of the extra coordinate $s$ on a single
quenched configuration at $\gbeta=5.85$
from the Columbia group\protect\cite{BOB}. 
The value of $M$ corresponds to 
a zero mode of $H_W(-M)$. The physical value is on the boundary, $s=N_s-1$.}
\label{pbp 90}
\end{figure}

Translationally invariant modes do exist in DWF simulations, however, so it is important
to understand their impact on the 4d physics.
In \FIG{pbp 90}, I show $\pbp$ as a function of the extra 5th 
coordinate. The plot is from the Columbia group\cite{BOB} and is 
for a single gauge configuration at $\gbeta=5.85$. 
The physical 4d value is at $s=N_s-1$. 
$M=1.45$, which corresponds to an exact zero of $H_W$. One sees
clearly the invariant mode, but its contribution to the physical value of 
$\pbp$ is only about one percent. Note, zero modes of $H_W(-M)$ are not
zeroes of DWF. In fact, these modes may be quite heavy; since
mixing between the modes on opposite walls is maximal in this case, 
the induced mass may be large.

\begin{figure}[hbt]
\vbox{\epsfxsize=3.0in \epsfbox{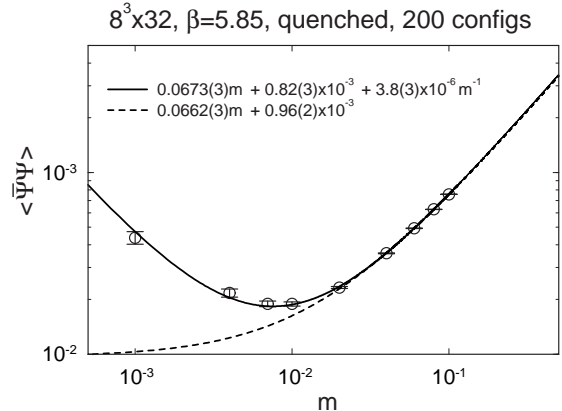} }
\vspace{-.25in}
\caption{$\langle\pbp\rangle$ from the Columbia group\protect\cite{FLEM} 
as a function of the quark mass. 
The presence of unsuppressed zero modes gives rise to a divergence as $m\to 0$.}
\label{pbp div}
\end{figure}

The Columbia group has also calculated $\langle\pbp\rangle$ on an ensemble of 200 gauge
fields at quenched $\gbeta=5.85$\cite{FLEM}. 
In \FIG{pbp div} a divergence as $m\to 0$ is clearly evident. The
divergence arises in the ensemble average from zero modes that are
unsuppressed by the (missing) fermion determinant. This has not been seen in lattice
simulations until now, presumably due to lattice artifacts. 
The coefficient of the $1/m$
term should decrease by $1/\sqrt{V}$. The results in \FIG{pbp div} are for
$V=8^3$. The $1/m$ term decreases by a factor of six on a 
$V=16^3$ lattice. The Columbia group is currently running on a still 
larger lattice to resolve this volume dependence.

\begin{figure}[hbt]
\vbox{\epsfxsize=3.0in \epsfbox{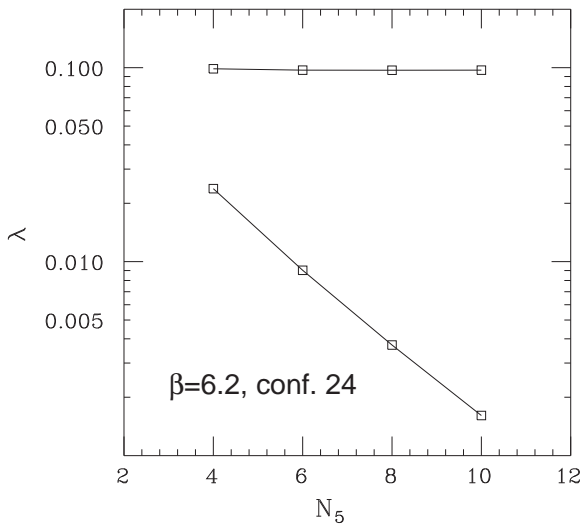}}
\vspace{-.25in}
\caption{ Eigenvalues of the hermitian DWF operator vs. $N_s$ on a
configuration with $Q_{top}=1$.
The plot is from the Argonne group\protect\cite{LAGAE}.}
\label{arg spect}
\end{figure}

\begin{figure}[htb]
\vbox{\epsfxsize=3.0in \epsfbox{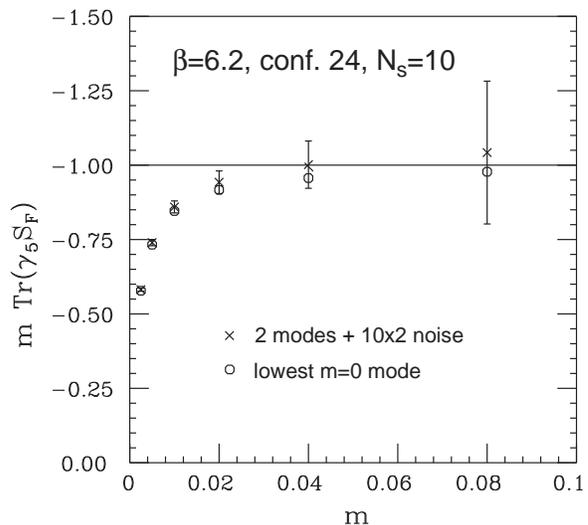}}
\vspace{-.25in}
\caption{The topological charge $Q_{top}$
from the Argonne group, \REF{LAGAE}, on a single configuration with
one anti-instanton.}
\label{top charge}
\end{figure}

Finally, the Argonne group has been calculating the low lying spectrum of 
$R \gamma_5 D_{dwf}$ in order to study the behavior of flavor singlet
correlation functions and the restoration of the $U(1)$ axial symmetry above
the chiral symmetry phase transition\cite{LAGAE}.
A sample of their results is shown in
\FIG{arg spect} for quenched $\gbeta=6.2$. Similar results hold at $\gbeta=6.0$.
Note that $N_s\le 10$ in the figure. The lowest eigenvalue is clearly exponential
in $N_s$ while the next lowest is $N_s$ independent.
For configurations with a net number of instantons, they find the
same  number of exponentially small eigenvalues,
in accord with the Atiyah-Singer Index theorem. 
They explicitly calculate the topological charge 
$Q_{top}\equiv m\int d^4x\bar q_x \gamma_5 q_x$
(\FIG{top charge}) and 
find that it is integer valued within small errors,
an impressive result considering the values
of $N_s$ used. The deviation of $Q_{top}$ near the origin is a finite $N_s$ effect.
Strictly speaking, the Index theorem for DWF is only exact
in the limit $N_s\to\infty$. Away from the continuum DWF(and the overlap)
do not have a unique index; it depends on $M$(see \REF{NARandVRAN}, for
example).

Thus, I conclude that the gap in $H_W(-M)$ is not actually closed for DWF simulations
at values of the coupling used in current simulations.
%for values of $M$ far from the $\kappa_c$ region, but below the threshold for $n_f=4$. 
Instead, DWF simply support a number of zero modes
corresponding to the {\it net} number of crossings in the spectrum of $H_W$ below
a chosen value of $M$, as they should. In the ensemble average and $V\to\infty$,
as long as the density of exact zero
modes of $H_W$ is vanishingly small for a particular $M$, then $M$ is suitable
for DWF simulations.

There is a trend for the suppression to weaken 
as $\gbeta$ decreases(see \FIG{anom ratio}). In
\REF{BS3} it was found that $N_s=18$ for $\gbeta=5.85$ was necessary for 
the pion mass 
to extrapolate to zero within statistical errors compared to $N_s=10$ at 6.0.
At 5.7 the Columbia group found that even for $N_s=48$, the pion mass has
a significant intercept at $m=0$\cite{BOB}. However, it is not clear
that this is entirely a finite $N_s$ effect since results from $N_s=24$ are
roughly the same. The intercept may be related to unsuppressed
zero modes from quenching. 
Our experience has been that at strong couplings there is a
minimum value of $N_s$ as well,
and tuning $M$ does not have a large effect.
The above behavior may be related to
an Aoki phase of Wilson fermions at strong coupling\cite{AOKI2}.

\subsection{Light quark masses and spectrum}
The quenched QCD pseudoscalar spectrum using DWF was first studied in 
Refs.\cite{BS1,BS2,BS3},
along with rough estimates for the vector channel. These measurements have also
been used to calculate the strange quark mass which was presented at
this conference\cite{WING}. The Columbia collaboration presented quenched results
for the light pseudoscalar, vector, and nucleon channels at $\gbeta=5.7$, as
well as the chiral condensate.

\begin{figure}[hbt]
\vbox{ \epsfxsize=3.0in \epsfbox[0 0 4096 4096]{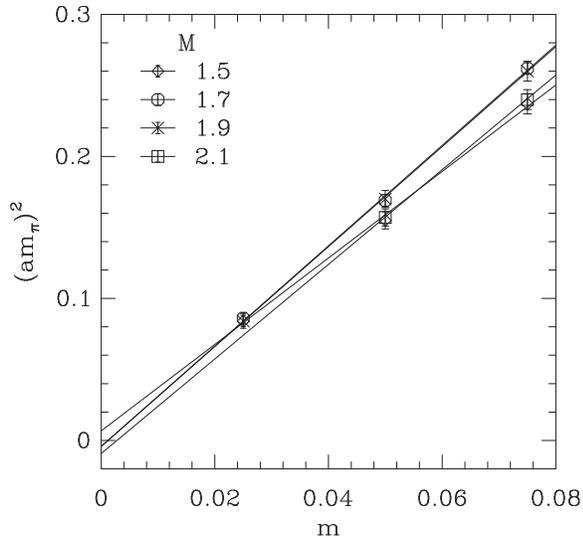} }
\vspace{-.25in}
\caption{The pion mass squared as a function of the quark mass.
$N_s=14$, $\beta=6.0$.}
\label{mpi2 vs m}
\end{figure}

\begin{figure}[hbt]
\vbox{ \epsfxsize=3.0in \epsfbox[0 0 4096 4096]{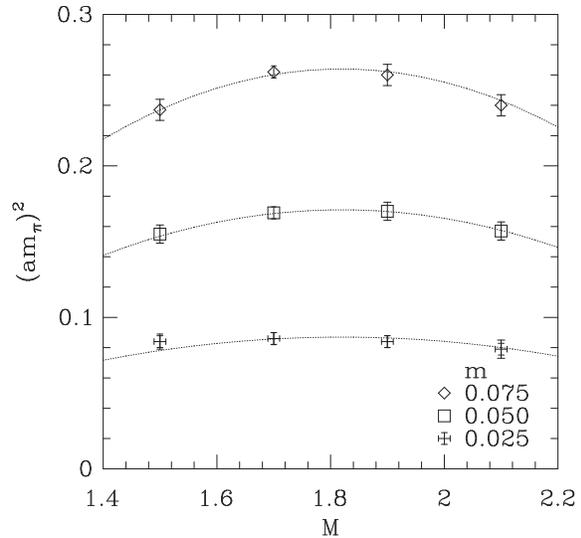} }
\vspace{-.25in}
\caption{The pion mass squared as a function of the domain wall height. The plot
is from \protect\REF{WING}. The dashed lines correspond to fits
to \protect\EQ{quark mass} using $\widetilde{M}$ and the nonperturbative value of $M_c$.}
\label{mpi2 vs M}
\end{figure}

In \FIG{mpi2 vs m} the pion mass squared is plotted as a function of
$m$ for $1.5\leq M \leq 2.1$ at $\gbeta=6.0$. Each curve linearly extrapolates
to zero, in accord with lowest order chiral perturbation theory.
\FIG{mpi2 vs M} shows the same data versus $M$.
$m_\pi^2$ has the characteristic quadratic shape expected from free field theory
but with a maximum that is shifted from its tree level value. From the figure
the shift is approximately 0.8, in good agreement with the estimates of $M_c$ above.
The dashed lines correspond to fits to \EQ{quark mass} with $M_c$ fixed to
the value corresponding to $\kappa_c$ for Wilson fermions(see \REF{WING} for
details).

The vector and nucleon masses from the Columbia group are shown in \FIG{mrho nuc
vs m}. The quenched coupling is 5.7 and the lattice volume is $8^3\times 32 \times 48$, so
these correspond to roughly the same physical volume as the above(in fact, the
rho mass scales within errors with the result in \REF{BS3}). From their
fits, $m_{nuc}/m_\rho\approx 1.4$ in the limit $m=0$ compared 
to roughly 1.5 calculated
using Kogut-Susskind quarks at the same physical volume\cite{GOTTLIEB}.
As mentioned above, the chiral symmetry breaking effects due to finite $N_s$ 
are larger at stronger coupling, so the Columbia group has used $N_s=48$. Even so,
the pion mass is not zero in the limit $m=0$, but corresponds to roughly 230 MeV.
%Whether or not this is a finite $N_s$ effect remains to be seen.
\begin{figure}[hbt]
\vbox{ \epsfxsize=3in \epsfbox{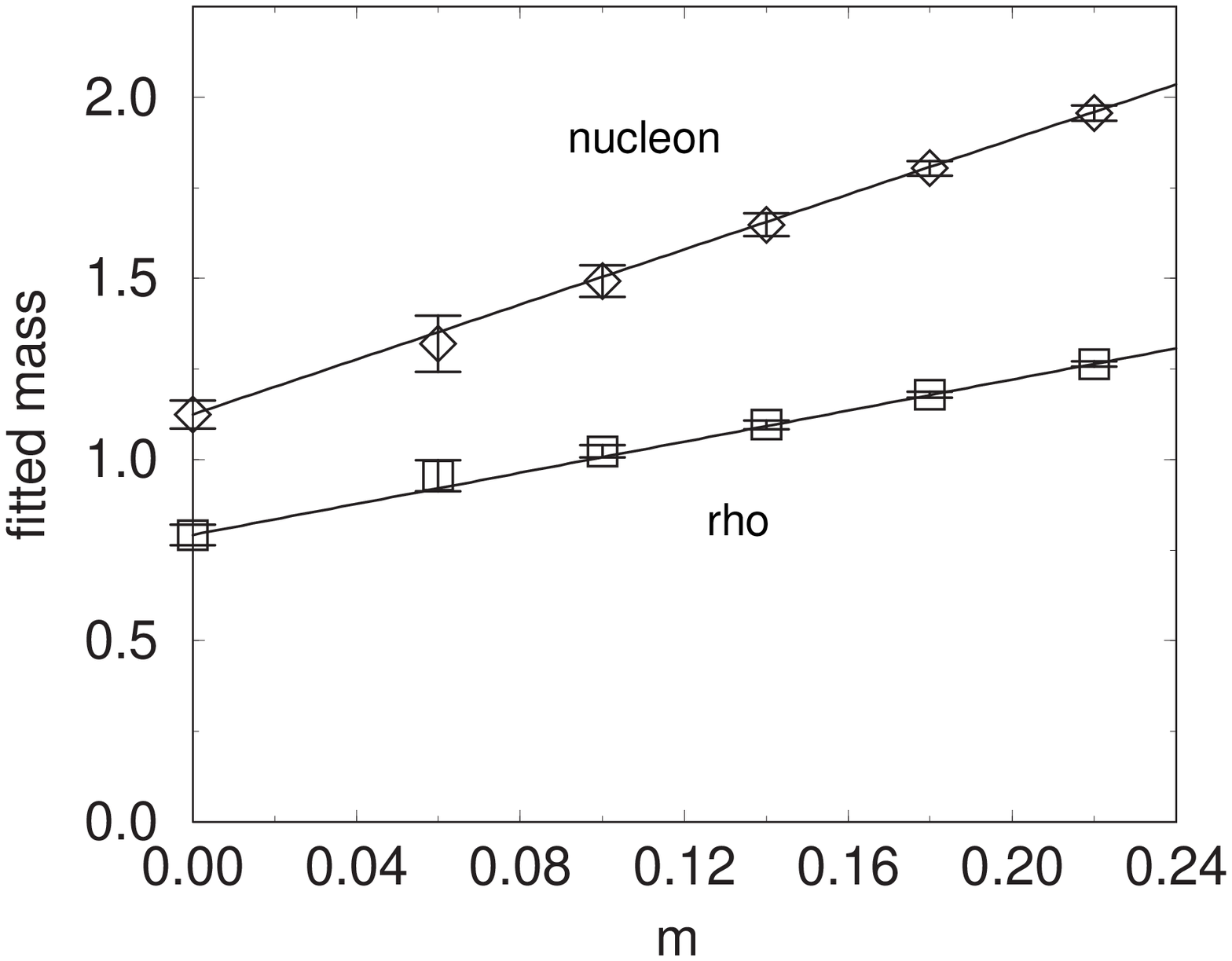} }
\vspace{-.25in}
\caption{The vector and nucleon masses as a function of $m$.
The plot is from \protect\REF{BOB}.}
\label{mrho nuc vs m}
\end{figure}

\begin{figure}[hbt]
\vbox{ \epsfxsize=3in \epsfbox{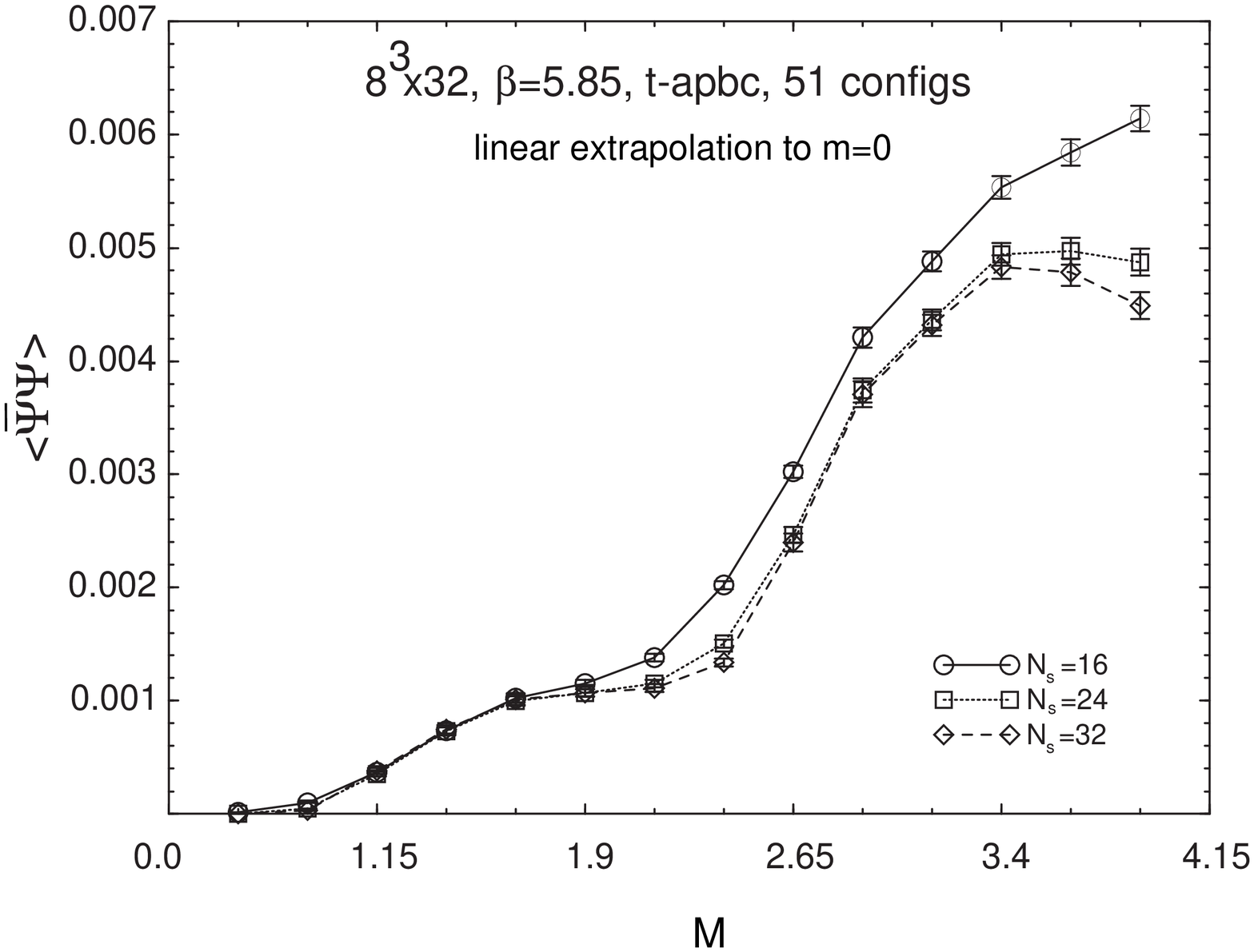} }
\vspace{-.25in}
\caption{$\langle\pbp\rangle$ as a function of $M$ from \protect\REF{BOB}.}
\label{pbp vs M}
\end{figure}

The Columbia group has calculated $\langle\pbp\rangle$ for several values of $M$ at 
quenched $\gbeta=5.85$
(\FIG{pbp vs M}) which yields a clear demonstration of the flavor structure of
DWF\cite{BOB}. Recall that in the free theory we 
had one flavor for $0< M < 2$, four
flavors for $2< M < 4$, and so on. A similar pattern emerges 
nonperturbatively, once again shifted to larger $M$. The light quark $\langle\pbp\rangle$ first
becomes 
non-zero at $M_c$, flattens out in the region $1.65\simle M \simle 2.15$, rises
rapidly again, and turns over at about 3.4. The first region corresponds to
one flavor and the
second, four. The ratio of $\langle\pbp\rangle$ in the two regions is numerically close to
four. In the free theory the regions are quadratic polynomials
with zeroes at $0, 2, 4, \ldots, 10$. Note that the results become $N_s$ 
independent at $N_s=24$.

\begin{figure}[hbt]
\vbox{ \epsfxsize=3.0in \epsfbox[0 0 4096 4096]{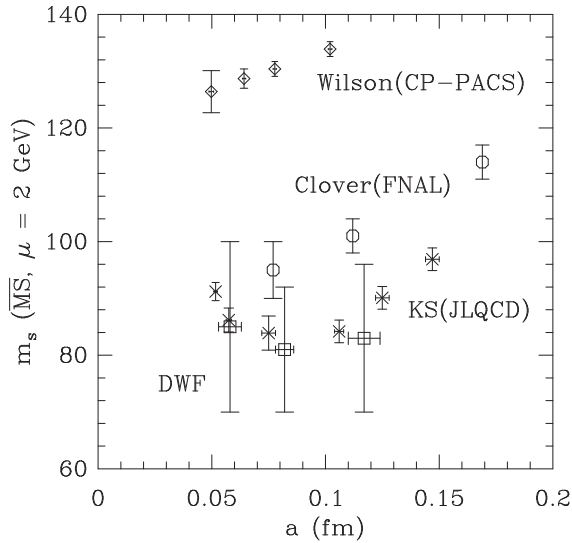} }
\vspace{-.25in}
\caption{The strange quark mass from \protect\REF{WING}
after matching to the $\overline {\rm MS}$ scheme.}
\label{strange quark mass}
\end{figure}

A first estimate of the (quenched) strange quark mass was presented
at the meeting\cite{WING}. The $\mu=2{\rm GeV}$ result
is compared to conventional determinations
in \FIG{strange quark mass}, where one sees rough agreement. The DWF result also 
appears to scale nicely with the lattice spacing, albeit within rather large
statistical errors. A weighted average gives $m_s=82(15)$MeV. 

To get $m_s$, several steps are needed.
First $m$ is tuned in the usual way to yield the kaon mass. 
To match the DWF result with a continuum scheme,
the one loop self-energy calculation of Aoki and Taniguchi was extended
to the massive case, $m\neq0$. 
One must determine a value for $\widetilde{M}=M-M_c$ to stick into the
one loop analog of \EQ{quark mass}. In the above it was determined
from $\kappa_c$. This method is not exact, even though it
is nonperturbative. That is because of the symmetry of the action under
the change $M\to 10-M$; $M=5$ is a fixed point so $M$ is not shifted uniformly.
While a more accurate determination of $\widetilde{M}$ is desirable, 
simply using $\kappa_c$ only introduces a small error in the final value of 
$\widetilde{M}(2-\widetilde{M})$\footnote{
I thank Y. Shamir for pointing this out.}. Ultimately, the best solution is
to calculate the quark mass renormalization $Z_m$ nonperturbatively
(using the method of \cite{SACH},
for instance), where this
extra factor is taken into account automatically.
\subsection{Weak interaction matrix elements}

The study of weak interaction matrix elements was one of the original
motivations for using DWF for QCD simulations\cite{BS1,BS2}. 
This is because the matrix elements
of many important weak operators between pseudoscalar states vanish linearly 
with the quark mass, which is a direct consequence of chiral 
symmetry\cite{BERNARD}. Ordinary Wilson quarks explicitly break chiral symmetry, so
the naive lattice transcription of these operators does not vanish in
the chiral limit. This is a well known, long standing problem. In principle,
the solution is to fine tune the continuum-like operators with others that transform
under different
chiral representations to restore the correct behavior in the chiral limit. 
Alternatively, one may use Kogut-Susskind quarks which do exhibit the correct
behavior since they maintain an exact abelian axial symmetry on the lattice. This
remnant of the continuum is enough to ensure the proper behavior. 
However, the price of Kogut-Susskind quarks is the breaking of the 
continuum flavor symmetries
which induces significant lattice artifacts. As we shall see, DWF seem 
to have significantly smaller discretization errors than the Kogut-Susskind variety.

\begin{figure}[hbt]
\vbox{ \epsfxsize=3.0in \epsfbox[0 0 4096 4096]{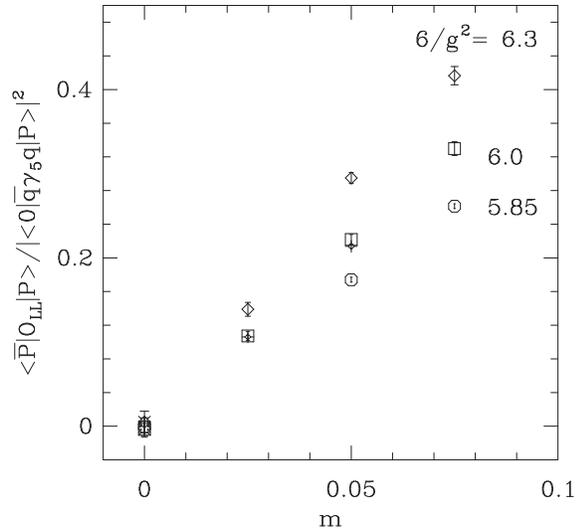} }
\vspace{-.25in}
\caption{The matrix element of $O_{LL}$ between degenerate
pseudoscalar states. It vanishes in the chiral limit, a stringent test
for the chiral symmetry properties of DWF. }
\label{mll}
\end{figure}

The vanishing of weak matrix elements in the chiral limit is a 
stringent test of the chiral symmetry of DWF since even with operator (un)mixing 
this is a difficult achievement
(see Refs.\cite{JLQCD-WIL,LELLOUCH} for recent results with nonperturbative and 
perturbative mixing, respectively). \FIG{mll} shows the
matrix element of the 
left-left weak operator ($O_{LL}=(\bar s \gamma_\mu (1-\gamma_5) d)^2$) 
that describes $K^0-\bar K^0$ mixing. For all three values of $\gbeta$, 
5.85, 6.0, and
6.3, it vanishes linearly with $m$, as it should
according to chiral perturbation theory\cite{BERNARD}.

\begin{figure}[hbt]
\vbox{ \epsfxsize=3.0in \epsfbox[0 0 4096 4096]{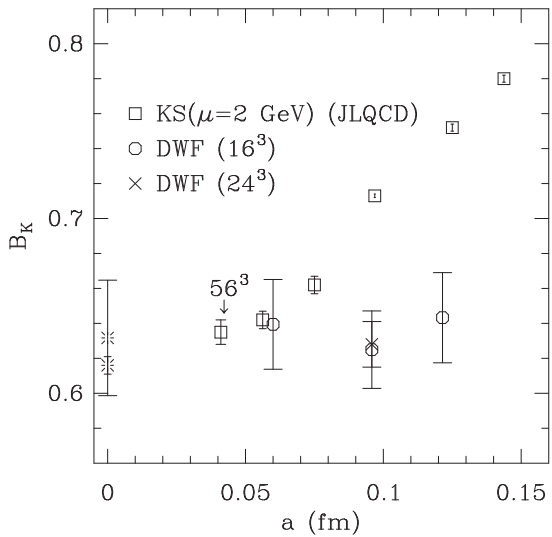} }
\vspace{-.25in}
\caption{The kaon $B$ parameter. The Kogut-Susskind result is
from \REF{JLQCD-BK}. DWF indicate improved scaling in this case.}
\label{bk}
\end{figure}

The matrix element of $O_{LL}$ is conventionally normalized to its value in the vacuum
saturation approximation which yields the $B$ parameter.
The kaon $B$ parameter, $B_K$, is
an important phenomenological parameter used to extract part of the CKM matrix
from experimental measurements, which in turn constrains the Standard Model. The
quenched DWF result is shown for several lattice spacings in \FIG{bk}. 
The impressive Kogut-Susskind result of \REF{JLQCD-BK} is shown for 
comparison. Since $O_{LL}$ is generated by the
operator product expansion, it is scale dependent, and therefore the result
at each coupling must be run to a common scale before the continuum limit is taken. 
In the continuum, the vacuum saturated value of 
$\langle K^0| O_{LL} | \bar K^0\rangle$
is proportional to the square of the axial current matrix element
which does not get renormalized since the axial current is partially conserved. 
As explained above, DWF have a partially conserved axial current, but this is not
the current that is needed for the vacuum saturation of $O_{LL}$ on the lattice 
since we have 
chosen to construct $O_{LL}$ from the quark operators defined in \EQ{quark ops}.
Therefore, the denominator in $B_K$ also receives a (finite) multiplicative
renormalization.
Since these renormalizations have not yet been done for DWF, the
most that can be said is that the result for $B_K$ is consistent with the
Kogut-Susskind result. Nevertheless, the indication from \FIG{bk} is that
discretization errors are significantly smaller in the DWF case. Needless to say,
it is a high 
priority to reduce the statistical errors and 
determine the renormalization factors in order to precisely determine the
lattice spacing dependence. It is important to point out that the lattice volumes
of the last two Kogut-Susskind points in \FIG{bk} are 
$56^3\times 96$ and $48^3\times 96$ and correspond to physical volumes of 
$2.3\rm fm^3$ and $2.8\rm fm^3$, respectively.
If the improved scaling of DWF holds up, then
the added cost of the extra dimension will be significantly offset by the
smaller $4d$ lattice volumes allowed by larger lattice spacings.

At $\gbeta=6.0$, there is also a value of $B_K$ on a $24^3\times 40$ lattice 
which indicates that finite volume effects are not large.

Encouraged by the results of our $B_K$ calculation, we have begun 
calculations of the $\Delta s=1$ effective weak Hamiltonian operators that
govern $K\to \pi \pi$ decays and are necessary for $\epsilon^\prime/\epsilon$ and
the $\Delta I=1/2$ rule, for example\cite{MARTINELLI}. Once again, in
the continuum, chiral symmetry restricts the behavior of these operators and
is therefore important for the lattice calculations as well. In fact, the
good chiral properties of DWF allow the use of the method of Bernard, 
\etal\cite{BERNARD} to calculate unphysical $K\to \pi$ amplitudes and 
relate them, using chiral perturbation theory,
to the physical $K\to\pi\pi$ amplitudes. The unphysical three point functions
are easier to calculate on the lattice.
Kilcup and Pekurovsky are using the same method with Kogut-Susskind quarks to 
calculate these amplitudes\cite{GREG}. There it is already clear that perturbation
theory fails for the renormalization of these operators, and a nonperturbative
method is needed. On the other hand,
since the one-loop renormalization of the DWF quark mass is
reasonable, it is still possible to renormalize the DWF operators 
perturbatively.

\begin{figure}[hbt]
\vbox{ \epsfxsize=3.0in \epsfbox[0 0 4096 4096]{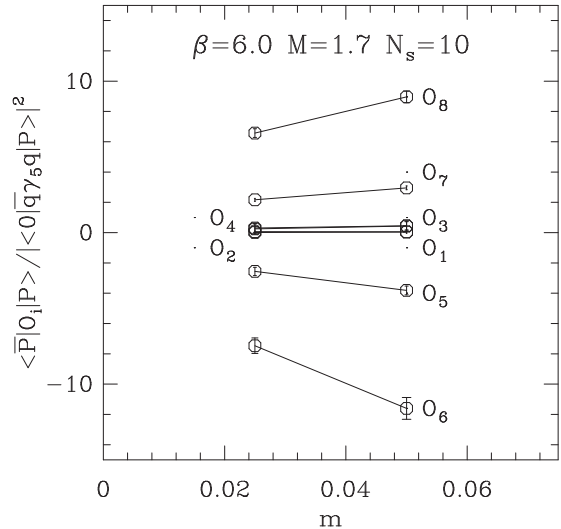} }
\vspace{-.25in}
\caption{$K\to\pi$ matrix elements of the $\Delta s =1$ effective weak
Hamiltonian.}
\label{delta s=1 ops}
\end{figure}

Our preliminary results on an ensemble of eighteen gauge configurations at
$\gbeta=6.0$ are shown in \FIG{delta s=1 ops}. The lattice volume is $24^3\times40
\times10$. As is well known,
these amplitudes are more difficult to calculate than $O_{LL}$ because of
the notorious ``eye" contractions which occur since a quark and an antiquark
with the same flavor in the operator can annihilate each other. In order to
efficiently average the operator over all spatial sites on a fixed time slice, 
one uses a random source(s) on the time slice.
The results shown in \FIG{delta s=1 ops}
are for a single random source in the middle of the lattice. It is therefore 
encouraging to see a signal at all. $O_1,\ldots,O_6$ which transform under 
(8,1) representations of $SU(3)_L\times SU(3)_R$ should vanish linearly
in the chiral limit. $O_5$ and $O_6$ do not, though they miss the origin by
less than two standard deviations. This could be due to the small
statistics of our ensemble, or that $N_s=10$ is not large enough. Also note
that there are only two quark masses, and the random source was different
for each. $O_{7,8}$ transform under (8,8) representations
so do not vanish in the chiral limit\cite{BERNARD}.
$O_{5,\ldots,8}$ are expected to be larger than $O_{1,\ldots,4}$ since they
are left-right operators which is the case here, and
the color mixed operators are roughly three times their unmixed counterparts,
as expected from the vacuum saturation approximation.

In order to carry the above through to the physical $K\to \pi\pi$ decays,
an important vacuum subtraction, $\langle K| O_i |0\rangle$
must still be made\cite{BERNARD}.

\subsection{Dynamical simulations and non-zero temperature}
The Columbia group has begun the first simulations using dynamical DWF
to study the two flavor QCD chiral symmetry restoration phase transition
at non-zero temperature. These simulations are obviously quite demanding.
Impressive preliminary results were presented at this 
conference\cite{VRANAS3}. They, as well as the Argonne group also presented 
quenched and semi-classical results related to the anomalous $U(1)$ axial 
symmetry\cite{KAEHLER,FLEM,LAGAE}. A detailed description of the Columbia
group's semi-classical results can be found in \REF{COL-SEMI-CLASS}.

\begin{figure}[hbt]
\vbox{ \epsfxsize=3.0in \epsfbox{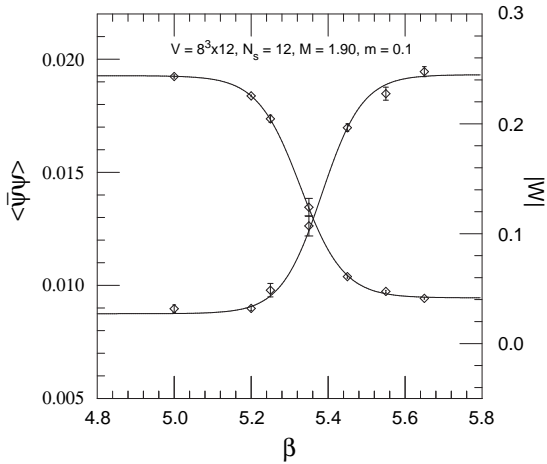} }
\vspace{-.25in}
\caption{$\langle\pbp\rangle$ and the Polyakov loop for 
$n_f=2$ from the Columbia collaboration\cite{VRANAS3}. Both exhibit a rapid 
crossover near $\beta=5.4$.}
\label{pbp wline vs beta}
\end{figure}

\begin{figure}[hbt]
\vbox{ \epsfxsize=3.0in \epsfbox{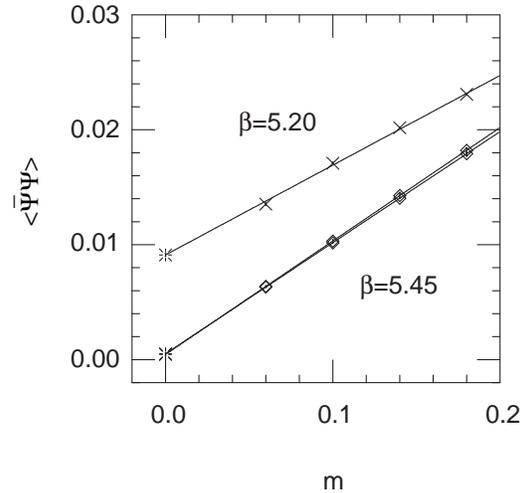} }
\vspace{-.25in}
\caption{$\langle\pbp\rangle$ as a function of $m$ for $\gbeta=5.20$(upper curve) and 
5.45(lower curves). Chiral symmetry is restored
above the phase transition. The two lower curves correspond to $8^3$ and $16^3$
lattices. $n_f=2$. The plot is from the Columbia group \REF{VRANAS3}.} 
\label{pbp vs m}
\end{figure}

Using DWF, the two flavor chiral phase transition in QCD appears qualitatively
similar
to the Kogut-Susskind result. In \FIG{pbp wline vs beta} there is a smooth
crossover in both the Polyakov loop and $\langle\pbp\rangle$ starting at $\gbeta\approx 5.2$
($N_\tau=4$). In the limit $m\to 0$,
below the transition $\langle\pbp\rangle$ is spontaneously broken while
above the transition chiral symmetry is restored 
(\FIG{pbp vs m}). 
The very small but statistically
significant non-zero intercept in \FIG{pbp vs m} is presumably due
to finite $N_s$ effects.
From \FIG{pbp vs m} ordinary finite volume effects are small as well. 
While these results are very encouraging, the quark masses are still too
heavy to discuss details like the order of the transition and critical exponents.
It should be stressed that for the first time 
the correct number of light pions is being simulated in dynamical studies
of the $n_f=2$ QCD chiral phase transition,
which may be crucial to the dynamics.


\begin{figure}[hbt]
\vbox{ \epsfxsize=3.0in \epsfbox{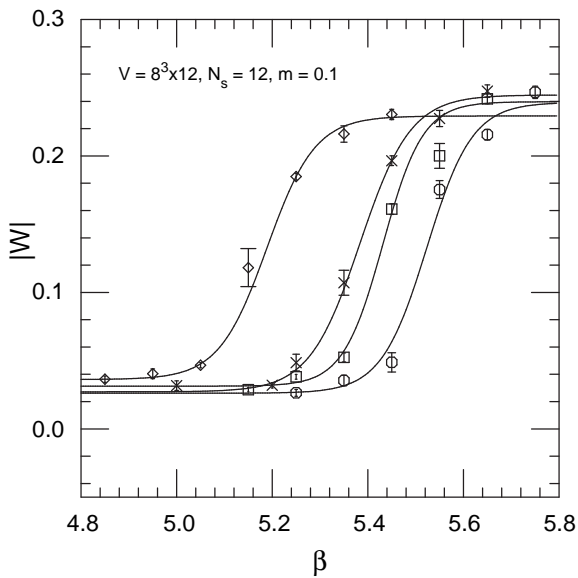} }
\vspace{-.25in}
\caption{The Polyakov loop from \REF{VRANAS3} for several values of
$M$. $n_f=2$. The pseudo-critical coupling depends on the choice of $M$.}
\label{poly vs M}
\vspace{-.25in}
\end{figure}

The pseudo-critical coupling is a function of $M$(\FIG{poly vs M}). 
There are two reasons for this: the quark mass depends on $M$ (\EQ{quark mass}),
and the number of flavors is a function of $M$. Below $M_c$ there are no light
states bound to the domain wall. At $M_c$ the domain wall 
states just begin to form so
$\beta_c$ starts at its quenched value. As $M$ increases, a full range of
momenta are available to the states on the walls and $\beta_c$ decreases to its
$n_f=2$ value(remember that the fermion determinant is $D^\dagger D$). 
As $M$ increases still further, four light quarks eventually
appear on the domain walls
so $\beta_c$ decreases to its $n_f=8$ value.

\begin{figure}[hbt]
\vbox{ \epsfxsize=3.0in \epsfbox{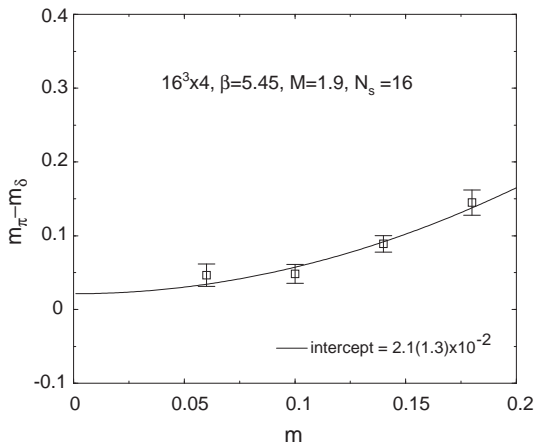} }
\vspace{-.25in}
\caption{The difference in screening masses above the chiral symmetry restoration
phase transition from \REF{VRANAS3}. $n_f=2$. A difference of zero at $m=0$ 
indicates restoration of the $U(1)_A$ symmetry. }
\label{mpi-mdelta}
\vspace{-.25in}
\end{figure}

Both the Argonne and Columbia groups have studied the restoration of the
$U(1)_A$ symmetry above the chiral phase transition.
As mentioned earlier, the Argonne group has measured the 
topological charge(see \FIG{top charge}) on quenched configurations, 
but have not yet published their results for flavor singlet
meson masses, correlators, or susceptibilities.
The Columbia group has measured $n_f=2$ screening masses for the pion and
the delta which should be degenerate if the $U(1)_A$ symmetry is restored.
The mass difference somewhat above the chiral transition is shown in 
\FIG{mpi-mdelta}. There is a small but significant value
at $m=0$. Note the data appears quadratic in $m$, as expected, due
to the restoration of chiral symmetry. There are no known lattice artifacts
present to obscure this picture as is the case for Kogut-Susskind quarks.

In general, dynamical simulations require a Pauli-Villars(PV) subtraction.
The action for the $N_s-1$
heavy states induced by the extra dimension 
dominates the total action in the limit $N_s\to\infty$\cite{NN1}. The form of
the subtraction is not unique; a second order operator is suggested 
in \REF{SHAMIR1}
while in \REF{VRANAS2,VRANAS3} a first order operator is used.
The latter version is vulnerable to the ``exceptional" configurations
discussed in section 3.1. That is, if $D_W(-M)$ supports an exact zero mode,
so does the first order operator, which has periodic instead of anti-periodic
boundary conditions in the extra dimension. This zero mode is not canceled
by a corresponding massless mode in the DWF operator. 
On the other hand, the authors of \REF{VRANAS3} have seen no evidence 
for these modes in their simulations. 
A method for suppressing such configurations
has been suggested in \REF{SHAMIR4}, which is described in the next section.
\section{TOPICS FOR FUTURE STUDY}

Beyond using DWF in their present form to do useful phenomenology, future 
studies will focus on improvements that reduce
the size of the extra dimension and/or increase the rate of 
exponential suppression. Several proposals have already 
been made\cite{BIET,KIK-YAM,SHAMIR4}.
The first proposal uses a hyper-cubic action instead of a Wilson
like action. The last two include an explicit lattice spacing in the extra dimension
which renders the DWF action positive definite for any gauge coupling. In \REF{SHAMIR4} it
is argued that this restricts eigenvalues of the transfer matrix on the unit circle 
to the real axis, reducing slowly decaying chiral symmetry violations. There, a modified
PV action is considered which suppresses the unwanted zero modes of $D_W(-M)$;
an extra term involving the 4d links is added to the PV action to do the job.
A side-effect is the small renormalization of the coupling constant.

An exciting development is the related derivation of
the overlap-Dirac operator by Neuberger\cite{NEU1}. 
This lattice discretization, which corresponds to a single massless Dirac particle,
is derived from
the overlap, but retains no reference to an extra flavor space or dimension. It is
completely 4d and has been shown explicitly to 
satisfy the Ginsparg-Wilson relation\cite{LUSCHER}.
Initial studies indicate the method may be feasible numerically\cite{CHIU,NEU3,SCRI2}, 
though its application to lattice QCD has not yet been demonstrated.

Finally, a new approach to lattice gauge theory is the quantum link model. 
DWF arise naturally in quantum link models since they are
constructed in five dimensions\cite{WIESE}.

\section{CONCLUSIONS}
I have reviewed the present status of DWF for vector gauge theories, in
particular QCD. We have seen that DWF with a modest number
of sites in the extra dimension maintain their good chiral properties,
which are exact in the limit $N_s\to\infty$.
The results show promising behavior in
studies where chiral symmetry is important such as weak matrix element phenomenology,
dynamical simulations of the chiral symmetry restoration phase transition, and
the relation between topology and fermionic zero modes. 
Indications of improved
scaling with DWF need to be verified since this may significantly
offset the cost of the added dimension.

I acknowledge many helpful discussions with M. Creutz,
R. Mawhinney, Y. Shamir, A. Soni, and M. Wingate. This work was supported by the
U.S. DOE, contract DE-AC02-98CH10886. Numerical simulations by the author were done on
the NERSC T3E.

\end{document}